\documentclass[runningheads,a4paper]{llncs}

\usepackage{amssymb}
\usepackage{graphicx}

\usepackage{url}
\urldef{\mailsa}\path|poli@jinr.ru, elena@jinr.ru, shukrinv@theor.jinr.ru|
\urldef{\mailsb}\path|anna.kramer, leonie.kunz, christine.reiss, nicole.sator,|
\urldef{\mailsc}\path|erika.siebert-cole, peter.strasser, lncs}@springer.com|
\newcommand{\keywords}[1]{\par\addvspace\baselineskip
\noindent\keywordname\enspace\ignorespaces#1}

\newcommand{\bs}{\begin{subequations}}
\newcommand{\es}{\end{subequations}}
\newcommand{\vp}{\varphi}
\RequirePackage{color}

\def \figr #1#2#3
{   \begin{figure}[ht]
        \centering\includegraphics[width=#2 \textwidth]{#1.eps}
        {\caption {#3}\label{#1}}
    \end{figure}
}   
\def \myfigures #1#2#3#4#5#6#7#8
{\begin{figure}[h!t]
    \centering\includegraphics[width=#2 \textwidth]{#1.eps}
        \hfill
    \centering\includegraphics[width=#6 \textwidth]{#5.eps}
        \parbox[t]{#4\textwidth}{\caption {#3}\label{#1}}
        \hfill
        \parbox[t]{#8\textwidth}{\caption {#7}\label{#5}}
\end{figure} }

\begin{document}
%
\title{Numerical investigation of the second harmonic effects in the LJJ}
\author{P.~Kh.~Atanasova,
\frame{T.~L.~Boyadjiev},
E.~V.~Zemlyanaya,
Yu.~M.~Shukrinov}
\authorrunning{Atanasova P.~Kh., Boyadjiev T.~L., Zemlyanaya E.~V., Shukrinov Yu.~M.}
\titlerunning{Numerical investigation of the second harmonic effects in the LJJ}
\institute{Joint Institute for Nuclear Research, Dubna 141980, Russia\\
\mailsa\\
\url{http://www.jinr.ru}
}
\maketitle
\begin{abstract}
We study the long Josephson junction (LJJ) model which takes into account the second harmonic of the Fourier
expansion of Josephson current. The sign of second harmonic is important for many physical applications. The
influence of the sign and value of the second harmonic on the magnetic flux distributions is investigated. At
each step of numerical continuation in parameters of the model, the corresponding nonlinear boundary problem is
solved on the basis of the continuous analog of Newton's method with the 4th order Numerov discretization
scheme. New solutions which do not exist in the traditional model have been found. The influence of the second
harmonic on stability of magnetic flux distributions for main solutions is investigated.
\end{abstract}
\section{Formulation of the problem}
Physical properties of magnetic flux in Josephson junctions (JJs) play important role in the modern superconducting
electronics. Tunnel SIS JJs  are known to be having the sinusoidal current
phase relation. However, the decrease of the barrier transparency in the SIS JJs leads
the deviations of the current-phase relation from the sinusoidal form
\cite{gki04}.

We study the static magnetic flux distributions in the long JJs taking into account the second harmonic in the Fourier-decomposition
of the Josephson current. This model is described by the double {sine}-Gordon equation (2SG) for
magnetic flux distribution in the static regime \cite{l85},\cite{htkt00},\cite{bk03}:
\begin{equation}\label{2sg}
    -\varphi\,'' + a_1 \sin \varphi + a_2 \sin 2\varphi -\gamma = 0,\quad x \in (-l;l),
\end{equation}
with the boundary conditions in the the following form
\begin{equation}\label{bound2sg}
    \varphi\,'(\pm l) = h_e.
\end{equation}
Here and below the prime means a derivative with respect to the coordinate $x$. The magnitude $\gamma$ is the external current, $l$ is the semilength of the junction,  $a_1$ and $a_2$ are parameters corresponding the  contribution of 1st and 2nd harmonic, respectively.
$h_e$ is external magnetic field. All the magnitudes are dimensionless.

The sign of the second harmonic depends on a physical application under study. It is important, in particular, in junctions like SNINS and SFIFS, where N is a normal metal and F is a weak metallic  ferromagnet \cite{ror01}. Interesting properties of long Josephson junctions with an arbitrarily
strong amplitude of the second harmonic in current phase relation were considered in \cite{goldobin07}.
We investigate the existence and stability magnetic flux distributions in dependence on the second harmonic contribution in both cases of negative and positive sign.

Stability analysis of $\vp(x,p)$ is based on numerical solution of the corresponding Sturm-Liouville problem \cite{galfil_84,pbvzpc07}:
\begin{equation}
\label{slp}
   -\psi\,'' + q(x)\psi = \lambda \psi, \quad
    \psi\,'(\pm l) = 0
\end{equation}
with a potential $q(x) = a_1 \cos \varphi + 2 a_2 \cos 2 \varphi$.

The minimal eigenvalue  $\lambda_0(p) > 0$ corresponds the stable solution. In case $\lambda_0(p) < 0$ solution $\vp(x,p)$ is unstable. The case $\lambda_0(p) = 0$ indicates the bifurcation with respect to one of parameters $p = (l,a_1,a_2,h_e,\gamma)$.
%
\section{Numerical scheme}
%
%
For numerical solution of the boundary problem (\ref{2sg}),(\ref{bound2sg}) we apply an iteration algorithm based on the continuous analog of Newton's method  (CANM) \cite{pbvzpc07}.
 Let an initial approximation for $\vp_0(x)$ be given. At $k\,^{th}$ step $(k = 1,2,, \ldots)$ we calculate:

1. Iteration correction $w_k(x)$ by solving linearized boundary  problem
\begin{equation} \label{namn_main}
 -w\,''_k + q_{k-1}(x) w_k = \varphi\,''_{k-1} -f_{k-1}(x)\,,
\end{equation}
\begin{equation} \label{namn_main_bc_left}
                w\,'_k(-l) = -\varphi\,'_{k-1}(-l) + h_e\,,\\
\end{equation}
\begin{equation} \label{namn_main_bc_right}
w\,'_k(l) = -\varphi\,'_{k-1}(l) + h_e\,,
\end{equation}
where $f(x) = a_1 \sin \vp + a_2 \sin 2 \vp - \gamma$.
\begin{table}[t]
\begin{center}
\begin{tabular}{|c|c|c|c|c|}
  \hline
      $x$ & $h=0.15625$ & $h = 0.078125$ & $h = 0.0390625$ & $\sigma \approx$ \\
  \hline
  $-5.00$ & 0.0539477770043 & 0.0539492562101 & 0.0539493470654 & $16.2809$ \\
  \hline
  $-3.75$ & 0.1018425717580 & 0.1018436848799  & 0.1018437558002  & $15.6954$ \\
  \hline
  $-2.50$ & 0.3299569440243 & 0.3299543520791 & 0.3299541941853 & $16.4157$ \\
  \hline
  $-1.25$ & 1.1169110352657 & 1.1169428501398 & 1.1169448259542 & $16.1022$ \\
  \hline
  $1.25$  & 5.1662742719134 & 5.1662424570391 & 5.1662404812249 & $16.1022$ \\
  \hline
  $2.50$  & 5.9532283631551 & 5.9532309551002 & 5.9532311129941 & $16.4157$ \\
  \hline
  $3.75$  & 6.1813427354214 & 6.1813416222995 & 6.1813415513793 & $15.6954$ \\
  \hline
  $5.00$  & 6.2292375301752 & 6.2292360509694 & 6.2292359601142 & $16.2809$ \\
  \hline
\end{tabular}
\end{center}
\caption{\label{t_err_runge_f} Values of function $\vp$ and quantities $\sigma$ (\ref{sigma}) in some points of the interval $[-l;l]$ for solution of kind  $\Phi^1$ at $2l = 10$, $\gamma = 0$, $h_e = 0$, $a_1 = 1$, $a_2 = 0$. }
\end{table}

2. Next approximation
$$ \varphi_{k}(x) = \varphi_{k-1}(x) + \tau_k\, w_{k}\,$$
where parameter $\tau_k$ is calculated by the Ermakov-Kalitkin formula \cite{ek_81}.

Further in order to simplify notations the iteration indices are omitted.

We introduce the grid $M_h = \{ x_i = -l+(i-1)h, i=\overline{{1,N}}, x_N = l, h = 2l/(N-1) \}$.
Numerov's  discrete approximation  \cite{numerov} of (\ref{namn_main})--(\ref{namn_main_bc_right}) yields the following
linear algebraic system with the three-diagonal structure at $i = \overline{3, N-2}$:
\[  -25w_1 + 48w_2 -36w_3 + 16w_4 -3w_5 = 12h (h_e - \vp\,'_1)  \]
\[  a_2 w_1 + b_2 w_2 + c_2 w_3 + d_2 w_4 + e_2 w_5 = r_2  \]
\[  a_iw_{i-1} + b_iw_i + c_iw_{i+1} = r_i\,,\quad i = \overline{3, N-2}\,, \]
\[  a_{N-1} w_{N} + b_{N-1} w_{N-1} + c_{N-1} w_{N-2} + d_{N-1} w_{N-3} + e_{N-1} w_{N-4} = r_{N-1}  \]
\[  25w_N - 48w_{N-1} + 36w_{N-2} - 16w_{N-3} + 3w_{N-4} = 12h (h_e - \vp\,'_N)  \]
where the coefficients  are determined by the following way
\[
a_2 = 1\,, \ b_2 = -2 - \frac{7h^2}{6}\, q_2\,, \ c_2 = 1 + \frac{5h^2}{12}\, q_3 \,, \ d_2 = -\frac{h^2}{3}\, q_4\,, \ e_2 = \frac{h^2}{12}\, q_5\,,
\]
\[
r_2 = \frac{h^2}{12}\,(14 f_2 - 5f_3 + 4f_4 - f_5) - \frac{h^2}{12}\,(14 \vp\,''_2 - 5\vp\,''_3 + 4\vp\,''_4 - \vp\,''_5)\,,
\]
\[
a_i = 1 - \frac{h^2}{12}\, q_{i-1}\,, \ b_i = -2 - \frac{5h^2}{6}\, q_i\,, \ c_i = 1 - \frac{h^2}{12}\, q_{i+1}\,,\ i = \overline{3, N-2}\,,
\]
\[
r_i = \frac{h^2}{12}\,(f_{i+1} + 10 f_i + f_{i_1}) - \frac{h^2}{12}\,(\vp\,''_{i+1} + 10 \vp\,''_i + \vp\,''_{i_1})\,, \ i = \overline{3, N-2}\,,
\]
\[
a_{N-1} = 1\,, \ b_{N-1} = -2 - \frac{7h^2}{6}\, q_{N-1}\,, \ c_{N-1} = 1 + \frac{5h^2}{12}\, q_{N-2} \,,
\]
\[
d_{N-1} = -\frac{h^2}{3}\, q_{N-3}\,, \ e_{N-1} = \frac{h^2}{12}\, q_{N-4}\,,
\]
\[
r_{N-1} = \frac{h^2}{12}\,(-f_{N-4} + 4 f_{N-3} - 5 f_{N-2} + 14 f_{N-1}) -
\]
\[
- \frac{h^2}{12}\,(-\vp\,'' + 4 \vp\,''_{N-3} - 5 \vp\,''_{N-2} + 14 \vp\,''_{N-1})\,,
\]
where  $\vp_i = \vp(x_i), \ q_i=q(x_i), \ f_i = f(x_i)$.

In order to test the accuracy order of the above numerical scheme we perform the calculations of (\ref{2sg}),(\ref{bound2sg}) at the sequence of uniform grids with steps $h$, $h/2$ and $h/4$ $(h=0.15625)$.  The results for solutions of the kind $\Phi^1$ are presented in the table~\ref{t_err_runge_f}. It is seen, the quantities $\sigma$ calculated by formula
\begin{equation}\label{sigma}
    \sigma(x_i) = \frac{\vp_{h}(x_i) - \vp_{h/2}(x_i)}{\vp_{h/2}(x_i) - \vp_{h/4}(x_i)}\,, \quad i = 1,2, \ldots, N,
\end{equation}
are close to $2^4$ that corresponds the theoretical accuracy order $O(h^4)$ of Numerov's approximation.

The  StLP (\ref{slp}) is approximated by the three-point finite-difference second order formulas \cite{bzh_60}.
The resulting algebraic eigenvalue problem is solved numerically with help of  a standard subroutine \cite{tridib}.

\section{Numerical results and conclusions} \label{numerical_results}
{\bf Trivial solutions.}
In the ``traditional'' case $a_2 = 0$ two trivial solutions $\varphi = 0$  and $\varphi = \pi$ (denoted by $M_{0}$ and $M_{\pi}$ respectively)  are known  at $\gamma = 0$ and $h_e = 0$. Accounting of the second harmonic $a_2 \sin 2\varphi$ leads to appearing two additional solutions  $\varphi = \pm \arccos (-a_1/2a_2)$ (denoted as $M_{\pm ac}$). The corresponding $\lambda_0$ as functions of 2SG-equation coefficients  have the form   $\lambda_0[M_0] = a_1 + 2 a_2$, $\lambda_0[M_{\pi}] = -a_1 + 2 a_2$ and $\lambda_0[M_{\pm ac}] = (a_1^2 - 4 a_2^2)/2 a_2$. The exponential stability of these constant solutions (CS) is determined by the signs of the parameters $a_1$ and $a_2$ and by its ratio $a_1/a_2$.
%

The full energy associated with the distribution of $\vp(x)$ is calculated by the formula \cite{pbvzpc07}
$$F(p) = \int \limits_{-l}^{l} \left[\frac{1}{2}\, \varphi\,'^2 + 1 - q(x) - \gamma \varphi \right]\,dx - h_e\Delta\varphi. $$
%
%

The full energy behavior in dependence on $a_2$ for considered distributions in the junction at $h_e=0$, $\gamma = 0$, $a_1=1$, $2l=10$
  is shown in Fig.\ \ref{fen_a1_1_2l10_he0_g0_comp_a2_cs_f1_upto1}.

  Stability properties of trivial solutions have been investigated in \cite{azbsh10}.

%
%
%
\figr{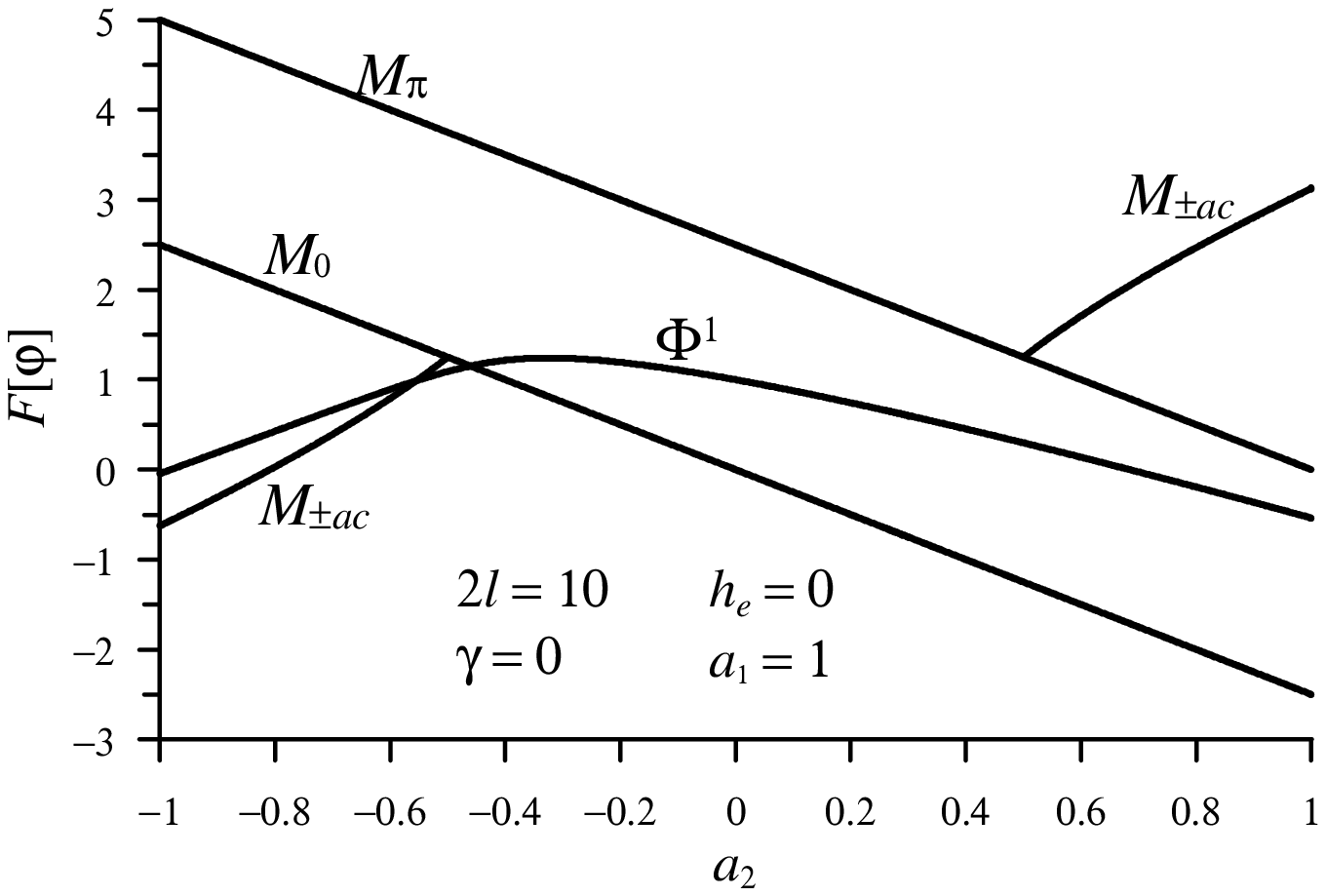}{0.6}{Full energy in dependence on $a_2 \in [-1;1]$  at $h_e = 0$, $\gamma = 0$ and $2l = 10$ for CS and $\Phi^1$.}
%
\figr{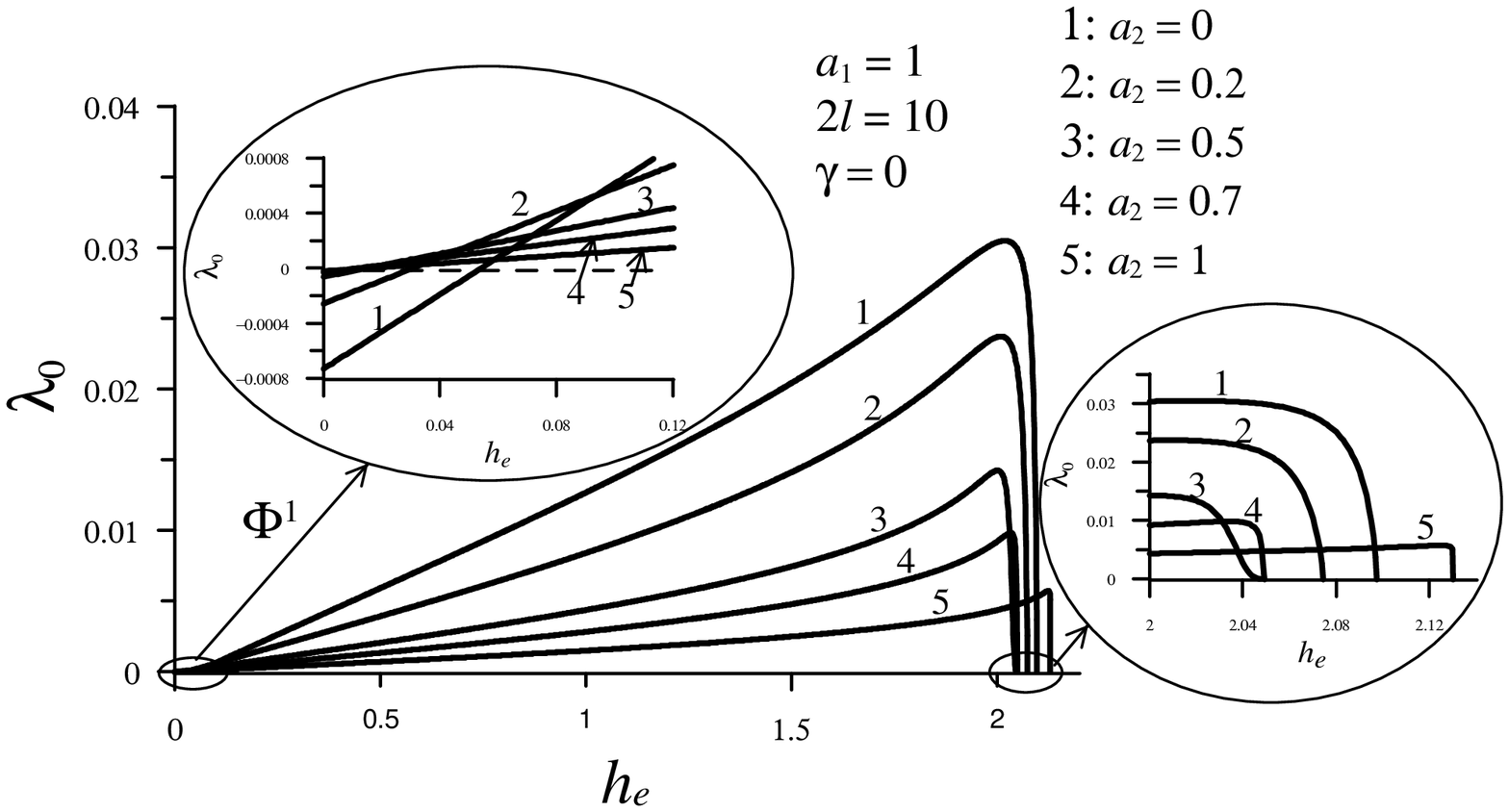}{1.}{Dependence $\lambda_0(h_e)$  for $\Phi^1$  at increase of $a_2 \in [0;1]$ and $a_1 = 1$,  $2l = 10$, $\gamma = 0$.}
%
\figr{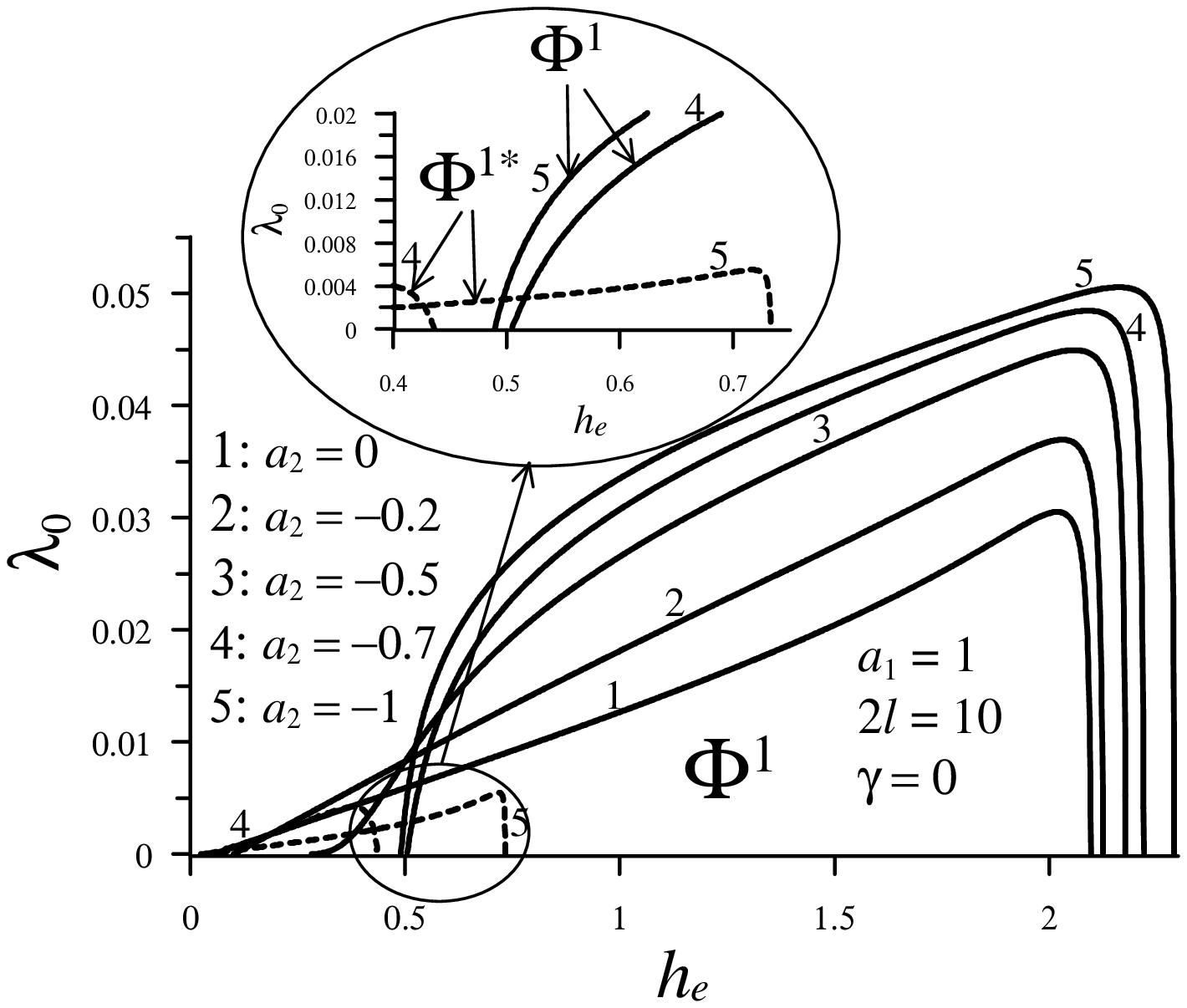}{0.8}{$\lambda_0(h_e)$ in dependence on $a_2 \in [-1;0]$ for $\Phi^1$ and $\Phi^{1*}$,
at $a_1 = 1$,  $2l = 10$, $\gamma = 0$.}
%
\figr{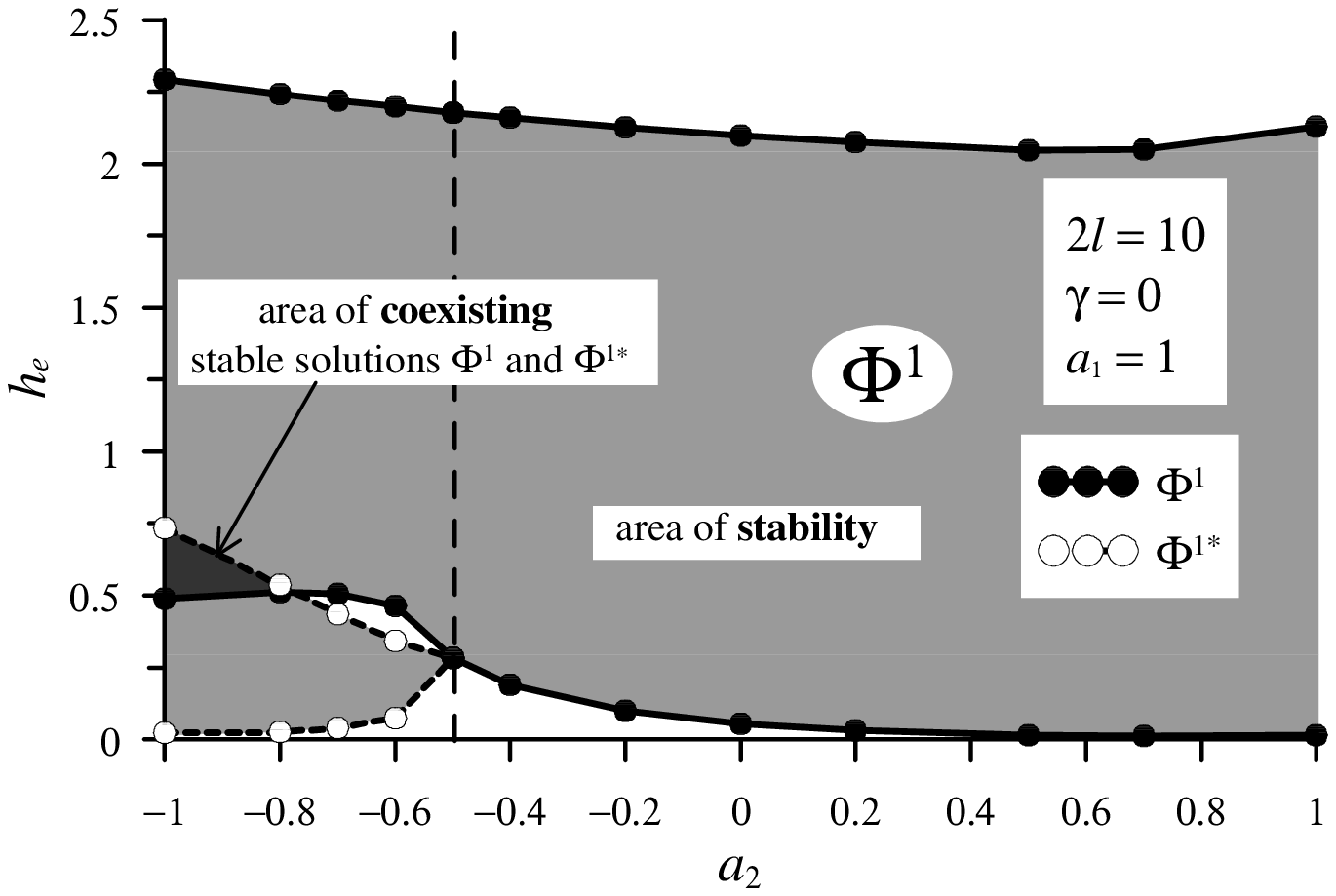}{0.7}{Bifurcation diagram of one-fluxon solutions at the plain of parameters $a_2$  and $h_e$. Here $a_1 = 1$,
$2l = 10$, $\gamma = 0$.}
%
%
%
%
{\bf Fluxon solutions.} The fluxons play an important role in the JJ physics. At small external fields $h_e$ such distributions are fluxon $\Phi^1$, antifluxon $\Phi^{-1}$ and their bound states $\Phi^1\Phi^{-1}$ and $\Phi^{-1}\Phi^1$ . As external magnetic field $h_e$ is growing, more complicated stable fluxon and bound states appear: $\Phi^{\pm n}$ and $\Phi^{\pm n}\Phi^{\mp n}$ $(n=1,2,3,...)$.

The energy of one-fluxon distribution $\Phi^1$ limits to unit $F(a_2 \to 0) \to 1$  which corresponds to an energy of a single fluxon $\Phi^1_\infty$ in a traditional ``infinite'' junction model  at $a_1=1$, $a_2 = 0$.
With change of $a_2$ the number of fluxons \cite{pbvzpc07}
$$ N(p) = \frac{1}{2l \pi} \int\limits_{-l}^l \varphi(x)\,dx,$$
corresponding to the distribution $\Phi^1$ is conserved i.e. $ {\partial N}/{\partial a_2 } = 0$. Here we have $N[\Phi^1] = 1$.

Let us discuss the main features of the dependence $\lambda_0(h_e)$ for one-fluxon state $\Phi^1$ in two intervals: $a_2 \in [0,1]$ and $a_2 \in [-1,0]$.

The change of the curve $\lambda_0(h_e)$ for one-fluxon state $\Phi^1$ when the parameter $a_2$ increases in the interval  $a_2 \in [0;1]$ is shown in Fig.~\ref{ev0_he_g0_a1_1_a2_0_02_05_07_1_f1_paper}. When $h_e = 0$, the state $\Phi^1$ remains unstable. With increase in $a_2$, $\lambda_0$ increases monotonically and tends to zero. With increase in magnetic field this solution becomes stable. The bifurcation point moves to the left with increase of parameter in the interval  $a_2 \in [0;0.7]$. At $a_2>0.7$ it goes to the right again. It follows a stability interval ending at $h_{cr} \approx 2$. The second bifurcation point also moves to the left at $a_2 \in [0; 0.5]$.
%
%
When $a_2$ is increased in $a_2 \in [0.5; 1]$, the bifurcation value $h_{cr}$ is also increased.

In the interval $a_2 \in [-1,0]$ we observe the following. When  $a_2$  increases in $(-0.5;0]$, the curve $\lambda_0(h_e)$ for $\Phi^1$ moves to the right (Fig.~\ref{ev0_he_g0_a1_1_a2_0_-02_-05_-07_-1_f1_paper}).
At  $a_2 < -0.5$ (case $a_2 = -0.7$ in Fig.~\ref{ev0_he_g0_a1_1_a2_0_-02_-05_-07_-1_f1_paper}), the curve corresponding to the stable solution $\Phi^1$ has two separate branches that are intersected at $a_2 \approx -0.8$). Here we observe a region along $h_e$, where two different stable one-fluxon  solutions
 (denoted by $\Phi^1$ and $\Phi^{1*}$) coexist, see Fig. \ref{bif_points_a2_f1_a1_1_pl}.
  %

%
Thus, we considered both positive and negative contributions of the second harmonic in 2GS equation. It is shown that its accounting leads appearing new constant solutions and changes the stability properties of the fluxon solutions.
Coexisting of two stable one-fluxon solutions requires further analysis and physical interpretation.

%
\textbf{Acknowledgements.}
Authors are thankfull to I.V.Puzynin and T.P. Puzynina for usefull discussions.
The work of  P.Kh.A. is partially supported in the frame of the Program for collaboration of JINR-Dubna and
Bulgarian scientific centers ``JINR -- Bulgaria''.
E.V.Z. was partially supported by  RFFI under grant 09-01-00770-a.
%

%

\begin{thebibliography}{99}
%
\bibitem{gki04}  Golubov, A.A.,  Kypriyanov, M.Yu.,  Il'ichev E.: The current-phase relation in Josephson junctions. Rev. Mod. Phys.  vol. 76, pp. 411--469 (2004)
%
\bibitem{l85}  Likharev, K.K.: Introduction in Josephson junction dynamics, M. Nauka, GRFML (in Russian) (1985)
%
\bibitem{htkt00}  Hatakenaka, N.,  Takayanag, H.,  Kasai, Yo.,  Tanda, S.: Double sine-Gordon fluxons in isolated long Josephson junction.  Physica B. vol. 284-288,  pp. 563-564 (2000)
%
\bibitem{bk03} Buzdin, A.,  Koshelev, A.E.:  Periodic alternating $0$-and $\pi$-junction structures as realization of $\varphi$-Josephson junctions. Phys. Rev. B.  vol. 67, p. 220504(R) (2003)
%
\bibitem{ror01} Ryazanov, V.V., Oboznov, V.A., Rusanov, A.Yu. et al.: Coupling of two superconductors through a ferromagnet: evidence for a pi junction. Phys. Rev. Lett.  vol. 36, pp. 2427--2430 (2001)
%
\bibitem{goldobin07}
Goldobin, E., Koelle, D., Kleiner, R., and Buzdin, A.: Josephson junctions with second harmonic in the current-phase relation:
Properties of  junctions. Phys. Rev. B. vol. 76, p. 224523 (2007)
%
\bibitem{galfil_84}  Galpern, Yu.S., Filippov, A.T.:  Joint solution states in inhomogeneous Josephson junctions.  Sov. Phys. JETP. vol. 59, p. 894 (in Russian) (1984)
%
\bibitem{pbvzpc07} Puzynin, I. V.,  Boyadzhiev, T. L.,  Vinitskii, S. I., Zemlyanaya, E. V.,  Puzynina, T. P., Chuluunbaatar, O.:   Methods of Computational Physics for Investigation of Models of Complex Physical Systems. Physics of Particles and Nuclei.  vol. 38, No. 1,  pp. 70–116 (2007)
%
%
%
\bibitem{ek_81} Ermakov, V.V, Kalitkin, N.N.: The optimal step and regularisation for Newton's method, USSR Comp.Phys.and Math.Phys. vol. 21, No. 2, p. 235 (in Russian) (1981)
%
%
\bibitem{numerov} Zemlyanaya, E.V.,  Puzynin, I.V.,  Puzynina, T.P.:
PROGS2H4 -- the software package for solving the boundary probem for
the system of differential equations.
JINR Comm. P11-97-414, Dubna,  18pp (in Russian) (1997)
%
\bibitem{bzh_60} Berezin, N.S., Zhidkov, E.P.: Numerical methods, M. Nauka, GRFML (in Russian) (1960)
%
\bibitem{tridib} TRIDIB -- translation of the ALGOL procedure BISECT,
    Num. Math. 9, 386-393(1967) by Barth, Martin, and Wilkinson.
    Handbook for Auto. Comp., vol.ii -- linear algebra, 249-256(1971).
%
\bibitem{azbsh10}Atanasova, P.Kh., Zemlyanaya, E.V., Boyadjiev, T.L., Shukrinov, Yu.M.: Numerical modeling of long Josephson junctions in the frame of double sin-Gordon equation. JINR Preprint  P11-2010-8, Dubna (2010); accepted to Journal of Mathematical modeling.
%
\end{thebibliography}
\end{document}